\documentclass[preprint,preprintnumbers,amsmath,amssymb]{revtex4}
\usepackage[dvips]{graphicx}
\usepackage{epsfig}
\usepackage{dcolumn}
\usepackage{bm}
\begin{document}
\title{Collective oscillations in two-dimensional Bose-Einstein condensate}
\author{Arup Banerjee \\
Laser Physics Division, Centre for Advanced Technology\\
Indore 452013, India\\}
\begin{abstract}
We study the effect of lower dimensional geometry on the frequencies of
the  collective oscillations of a Bose-Einstein
condensate confined in a trap. To study the effect of
two dimensional geometry we consider a pancake-shaped condensate confined in a harmonic trap and employ various models for the coupling constant
depending on the thickness of the condensate relative to the the
value of the scattering length. These models correspond to different scattering regimes ranging from quasi-three dimensional to strictly  two dimensional regimes. Using these models for the coupling parameter and sum rule approach of the many-body response
theory we derive analytical expressions for the frequencies of the
monopole and the quadrupole modes. We show that the frequencies of
monopole mode of the collective oscillations are significantly
altered by the reduced dimensionality and also study the evolution of the frequencies as the system make transition from one regime to another.
\end{abstract}

 \maketitle Recently, several theoretical and experimental
studies devoted to the influence of dimensionality on the
properties of a Bose-Einstein condensate in a harmonic trap have
been reported in the literature
\cite{baganto,mulin,ho,petrov1,petrov2,shevchenko,kolomeisky,lieb,lee,salasnich,tkghosh,tantar1,tantar2,kamchatnov,gorlitz,arimondo}.
The reduction of dimensionality affects the physical properties of
the condensates resulting in very different features from their
three dimensional (3D) counterparts. In current experiments on
Bose-Einstein condensates with alkali atoms confined in a magnetic
trap the anisotropy parameter $\lambda$ (defined as the ratio
between the frequencies of the trap in the z- and the
transverse directions) may be varied to achieve condensates with
special properties which are characteristic of the low
dimensionality. For example, by making the anisotropy parameter
much larger than unity ($\lambda >>1$) a flatter and flatter
(pancake-shaped) condensate can be produced. Such a pancake-shaped
condensate is expected to exhibit special features of two
dimensional (2D) condensate. It is well known that in the 2D case
the scattering properties are very different as compared to the 3D
case and this in turn lead to a significant modification of the
boson-boson coupling constant. For example, the boson-boson
coupling constant in 2D limit becomes density dependent even at a
low density and zero temperature. In contrast to this for 3D case,
to the lowest order in the density the interactions are described
by constant coupling strength and deviations occur only when
quantum depletion and finite temperature effects are taken into
considerations \cite{burnett}.

In a pancake shaped 3D condensate as the anisotropy parameter is
increased the physical properties of the condensate first change
due to the modified shape of the confinement \cite{salasnich} and
then also due to the alteration of the scattering properties. With
respect to the scattering properties of a flat pancake-shaped
condensate three regimes can be identified which are characterized
by different expressions for the coupling constant \cite{tantar1}.
When the linear dimension of the condensate along z-axis given by
$a_{z} = (\hbar/m\omega_{z})^{1/2}$ (where $\omega_{z}$ is the
z-component of the angular frequency of the trapping potential and
$m$ is the mass of the trapped atoms) is still much larger than
the 3D scattering length $a$ ($a_{z}>>a$), the collisions still
take place in three dimensions and this is referred to as quasi-3D
(Q3D) regime. On further increasing the anisotropy parameter the
condensate gets more tightly confined along the z-axis but the
assumption that the scattering is unaffected begins to break down
when $a\approx a_{z}$ and the condensate is said to be in quasi-2D
(Q2D) regime. A fully 2D condensate is achieved when $a_{z}<<a$ so
that the collisions are restricted only in the transverse
$\{x,y\}$ plane. It is natural to expect that as the condensate
evolves from a fully 3D to a fully 2D regime its static and
dynamic properties undergo dramatic changes. In Ref.
\cite{tantar1} the evolution of the density profile of a
condensate has been investigated as the system crosses from the 3D
to the 2D regime. They have found that the widths of the density
distribution crucially depend on the collisional properties or the
boson-boson interaction parameters in the different regimes.

The main aim of this letter  is to study the effect of lower
dimensionality on the the frequencies of the collective
oscillations of a condensate. In particular we calculate the
collective oscillation frequencies of the quadrupole and monopole
modes of a flat pancake-shaped condensate and study how these
frequencies evolve as the system undergo transitions from the Q3D
to the Q2D and to a strictly 2D regime. For this purpose we make
use of the sum rule approach of many-body response theory
\cite{bohigas,lipparini}. By employing this method we derive
analytical expressions for the frequencies of the quadrupole and
the monopole modes. In the following we first derive these
expressions and then discuss the results.

We consider a dilute condensate with $N$ bosons  confined in an
anisotropic (pancake shaped) harmonic trap characterized by the
frequencies $\omega_{\bot}$ and $\omega_{z} =
\lambda\omega_{\bot}$ with the anisotropy parameter $\lambda$
being much larger than unity. Within the density functional
theory, the ground properties of a condensate can be completely
described by the ground state condensate density $\rho ({\bf r})$
in $\{x,y\}$ plane. The ground state density of the condensate can
be determined by minimizing the local density energy functional
\begin{equation}
E[\rho ] = \int d^{2}{\bf r}\left
[-\frac{\hbar^{2}}{2m}|{\bf\nabla}\sqrt{\rho({\bf r})}|^{2} +
v_{ext}({\bf r})\rho ({\bf r}) + \epsilon (\rho )\rho ({\bf
r})\right ], \label{energyfunc}
\end{equation}
where $v_{ext}({\bf r})$ is the external harmonic potential in the
transverse direction given by
\begin{equation}
v_{ext}({\bf r}) = \frac{1}{2}m\omega_{\bot}^{2}\left (x^{2} +
y^{2}\right ). \label{harmonicpot}
\end{equation}
In the above equation (\ref{energyfunc}) the first and the third
terms represent the kinetic energy and the energy due to the
interatomic interaction within local density approximation (LDA)
respectively. Within this approximation the interaction energy per
particle $\epsilon(\rho)$ is given by
\begin{equation}
\epsilon(\rho) = \frac{g}{2}\rho({\bf r}).
\end{equation}
Where $g$ is the coupling constant whose form depends on the
collisional properties of the condensate. For example, $g$ is
independent of the density for the 3D case, on the other hand in
the purely 2D regime it depends logarithimically on the density.
In the following we briefly describe the models for the coupling
constant $g$ valid in the different collisional regimes.

For the 3D system the coupling parameter $g$ which is a constant
completely determined by the s-wave scattering length $a$  and it
is given by
\begin{equation}
g = \frac{4\pi\hbar^{2}}{m} a.
\end{equation}
When the linear dimension $a_{z}$ of the condensate along the z-direction
is much larger the 3D scattering length $a$ ($a_{z}>>a$), the collisions still
take place in three dimensions. Under this condition the effective coupling
constant $g_{Q3D}$ which includes the effects of reduced dimensionality
only is given by \cite{ho,salasnich}
\begin{equation}
g_{Q3D} = 2\sqrt{2\pi}\frac{\hbar^{2}a}{ma_{z}} \label{gq3d}
\end{equation}
On further increasing the anisotropy and and $a_{z}$ becoming
comparable to $a$ ($a_{z}\approx a$), the collisions start getting
affected by the tight confinement along the z-direction. Under such a condition
the condensate is said to be in Q2D regime. The coupling
constant in this regime is given by
\begin{equation}
g_{Q2D} = \frac{2\sqrt{2\pi}\frac{\hbar^{2}a}{ma_{z}}}{1 +
\frac{a}{\sqrt{2\pi}a_{z}}|ln\left (2(2\pi)^{3/2}\rho({\bf
r})aa_{z}\right )|} \label{gq2d}
\end{equation}
Here $\rho({\bf r})$ is the ground state density of the condensate. The above
expression was originally derived by Petrov et al.
\cite{petrov1,petrov2} by studying the scattering properties of a
bosonic system which is trapped harmonically trapped in
the z-direction and uniform in the $\{x,y\}$ plane. A similar
expression was later derived by Lee et al. \cite{lee} also by
employing many-body T-matrix approach. It is important to note
here that in the Q2D regime the coupling constant  becomes dependent on the density
in accordance with the behaviour of collisions in two dimensions.
Finally, when $a_{z}$ becomes much smaller than $a$ ($a_{z}<< a$),
the collisions can be safely assumed to be taking place in two
dimensions resulting in a 2D condensate. The coupling constant in
2D regime is given by
\begin{equation}
g_{2D} = \frac{4\pi\hbar^{2}}{m}\frac{1}{|ln\rho({\bf r})a^{2}|}
\label{g2d}
\end{equation}
Notice that the expression for $g_{2D}$ also depends on the
density $\rho({\bf r})$ but the information about the confinement
direction is absent as it corresponds to the purely 2D case. The expression for
$g_{2D}$ was derived in Ref. \cite{schick} for a homogenous bose gas of hard disc.
This form of $g_{2D}$ has been employed to study the properties cofined bosons in two dimensions \cite{shevchenko,kolomeisky} and the rigorous justification for this use was provided by Lieb et al. \cite{lieb}.

Having described the different models for the coupling constant, we now briefly discuss
the method for obtaining the frequencies of the monopole and the quadrupole modes
of collective oscillations. As mentioned earlier for this purpose
we employ the sum-rule approach of the many-body response theory.
The most important advantage of this method is that the
calculation of frequencies requires the knowledge of the
ground-state wave function (or the corresponding ground-state
density) of many body system only. In accordance to the basic results of
the sum-rule approach \cite{bohigas,lipparini} the upper bound of
the lowest excitation energy is given by
\begin{equation}
\hbar\Omega_{ex} = \sqrt{\frac{m_{3}}{m_{1}}}
\label{ubfreq}
\end{equation}
where
\begin{equation}
m_{p} = \sum_{n}|\langle 0|F|n\rangle |^{2}\left
(\hbar\omega_{0n}\right )^{p}
\end{equation}
is the $p$-th order moment of the excitation energy
$\hbar\omega_{0n}$ associated with the excitation operator $F$ and
$\Omega_{ex}$ is the frequency excitation. Here
$\hbar\omega_{n0}=E_{n}-E_{0}$ is the excitation energy of
eigenstate $|n\rangle$ of the Hamiltonian $H$ of the system. The
upper bound given by Eq. (\ref{ubfreq}) is close to the exact
lowest excited state when this state is highly collective, that
is, when the oscillator strength is almost exhausted by a single
mode. This condition is satisfied by the trapped bosons in most of
the cases. Moreover, Eq. (\ref{ubfreq}) can be used for
computation of the excitation energies by exploiting the fact that
the moments $m_{1}$ and $m_{3}$ can be expressed as expectation
values of the commutators between $F$ and $H$ in the ground state
$|0\rangle$ \cite{bohigas,lipparini}:
\begin{eqnarray}
m_{1} & = & \frac{1}{2}\langle 0|\left [F^{\dagger},\left [H,F\right ]\right ]|0\rangle , \nonumber \\
m_{3} & = & \frac{1}{2}\langle 0|\left [\left [F^{\dagger},H\right ],\left [\left [H,[H,F\right ]\right ]\right ]
|0\rangle .
\label{7}
\end{eqnarray}
For the purpose of calculation of $m_{1}$ and $m_{3}$ as given by the above equation we need to first choose an appropriate excitation opertaor $F$. Following Ref. \cite{tkghosh} the excitation opertaor $F$ is written as
\begin{equation}
F = x^{2} + \alpha y^{2}
\end{equation}
with $\alpha = 1$ and $\alpha = -1$ for the monopole and the
quadrupole modes respectively. By using the energy functional given by
Eq. (\ref{energyfunc}) along with the expression for $\epsilon
(\rho)$ we find after some tedious although straightforward algebra
following expressions for the frequencies of the quadrupole
\begin{equation}
\Omega_{q} = \sqrt{2}\left (1 + \frac{T}{U}\right )^{1/2},
\label{freqq}
\end{equation}
and the monopole
 \begin{equation}
\Omega_{m} = \sqrt{2}\left (1 + \frac{T}{U} + \frac{Y_{int}}{U}\right )^{1/2},
\label{freqm}
\end{equation}
modes. In the above equations $T$ and $U$ denote the kinetic and the harmonic confinement energies and they are given by
\begin{eqnarray}
T & = & \frac{\hbar^{2}}{2m}\int d^{2}{\bf r}|{\bf\nabla}\sqrt{\rho({\bf r})}|^{2} \nonumber \\
U & = & \frac{1}{2m}\omega_{\bot}^{2}\int d^{2}{\bf r} \left ( x^{2} + y^{2}\right)
\rho({\bf r})
\end{eqnarray}
On the other hand $Y_{int}$ arises from the interaction energy (third term) term of Eq. (\ref{energyfunc}).  As a result of this the values of $Y_{int}$ depend on the model of the coupling constant.  The expressions for $Y_{int}$ (in the unit of $N\hbar\omega_{\bot}$) corresponding to the three regimes are given by
\begin{eqnarray}
Y_{int}^{Q3D} & = & E_{int}^{Q3D} \nonumber \\
Y_{int}^{Q2D} & = & \left (E_{int}^{Q2D} + 2kI_{1}^{Q2D} + 2k^{2}I_{2}^{Q2D}\right ) \nonumber \\
Y_{int}^{2D} & = & \left (E_{int}^{2D} + 2I_{1}^{2D} - 2I_{2}^{2D}\right )
\label{eqyint}
\end{eqnarray}
with $k = \tilde{a}/\sqrt{\pi}$ and the dimensionless interaction parameter $\tilde{a}=a/a_{z}$. The general expression for the interaction energy
$E_{int}^{i}$ ($i = Q3D, Q2D, 2D$) is given by
\begin{equation}
E_{int}^{i} = \int d^{2}{\bf r}\frac{g_{i}}{2}\rho^{2}({\bf r})
\end{equation}
By using coupling constant given by Eqs. (\ref{gq3d})- (\ref{g2d})in the above integral we obtain the interaction energies in the corresponding regime.  On the other hand, the expressions for other terms appearing in Eq. (\ref{eqyint}) can be written as
\begin{eqnarray}
I_{1}^{Q2D} & = & \sqrt{2\pi}N\tilde{a}\int\frac{d{\bf
r}^{2}\rho^{2}({\bf r})}{1 + k |ln\left (2(2\pi)^{3/2}N\rho({\bf
r})\frac{\tilde{a}}{\lambda}\right )|^{2}}\nonumber \\
I_{2}^{Q2D} & = & \sqrt{2\pi}N\tilde{a}\int\frac{d{\bf
r}^{2}\rho^{2}({\bf r})}{1 + k |ln\left (2(2\pi)^{3/2}N\rho({\bf
r})\frac{\tilde{a}}{\lambda}\right )|^{3}}\nonumber \\
I_{1}^{2D} & = & 2\pi N\int\frac{d{\bf r}^{2}\rho^{2}({\bf r})}{
|ln\left (N\rho({\bf
r})\tilde{a}^{2}\right )|^{2}}\nonumber \\
I_{2}^{2D} & = & 2\pi N\int\frac{d{\bf r}^{2}\rho^{2}({\bf r})}{
|ln\left (N\rho({\bf r})\tilde{a}^{2}\right )|^{3}} \label{eqi}
\end{eqnarray}
The density appearing in the above equation (Eq. (\ref{eqi})) are
normalized to unity and notice that the anisotropy parameter
$\lambda$ is explicitly appearing in the integrals for the Q2D case. Before proceeding further we note that the expressions given by Eqs. (\ref{freqq}) and (\ref{freqm}) for the Q3D case match with the corresponding results of Ref. \cite{tkghosh}.

It is evident from Eq. (\ref{freqq}) that the frequency of the
qudrupole mode is not explicitly dependent on the interaction
energy and it is true in all the three regimes considered in this
paper . We wish to note here that the Eq. (\ref{freqq}) is
identical to the expression for the frequency of the quadrupole
mode of a 3D condensate with constant coupling parameter \cite{stringari}.
In the absence of inter particle interactions we have $T = U$ and this leads to
harmonic oscillator result $\Omega_{q} = 2$. On the other hand, when
interaction energy is much larger than the kinetic energy so that
the kinetic energy can be neglected ( Thomas-Fermi approximation)
we get $\Omega_{q} = \sqrt{2}$. In the general case one needs to
know the value of kinetic energy of the ground state for accurate
estimation of the frequency of quadrupole mode.

Now we focus our attention on the frequency of the monopole mode
of the collective oscillations. In contrast to the quadrupole case
the frequencies of monopole mode are explicitly dependent on the
interaction energy. Therefore, the frequencies of the monopole mode
will be affected by different models of the coupling
constant. To illustrate the dependence of the interaction energy on the frequencies
more clearly we substitute the virial relations associated with the three models
\begin{eqnarray}
T - U + E_{int}^{Q3D} & = & 0 \nonumber \\
T - U + E_{int}^{Q2D} + kI_{1}^{Q2D}& = & 0 \nonumber \\
T - U + E_{int}^{2D} + I_{1}^{2D}& = & 0 \label{eqvirial}
\end{eqnarray}
in the respective expressions in Eq. (\ref{freqm}) to obtain
\begin{eqnarray}
\Omega_{m}^{Q3D} & = & 2 \nonumber \\
\Omega_{m}^{Q2D} & = & 2\left (1 + k\frac{I_{1}^{Q2D}}{2U} +
k^{2}\frac{I_{2}^{Q2D}}{U}\right )^{1/2}
\nonumber \\
\Omega_{m}^{2D} & = & 2\left (1 + \frac{I_{1}^{Q2D}}{2U} -
\frac{I_{2}^{Q2D}}{U}\right )^{1/2} \label{eqmonopole}
\end{eqnarray}
The first of the above equation shows that unlike the Q2D and the
2D cases the frequency of the monopole mode in
the Q3D regime is independent of the coupling constant. The frequency of the monopole mode in the Q3D regime is given by the frequency of a 2D harmonic ocsillator.
It is important to point out that in contrast to the Q3D case the frequency
of the monopole mode of a 3D condensate is
explicitly dependent on the interaction strength \cite{stringari}.

Now we turn to the detail study of the frequencies of the monopole mode for
the Q2D and the 2D models of the coupling constant. To this end we
first need to evaluate the integrals given in the Eq. (\ref{eqi}).
For this purpose we employ the ground state densities $\rho ({\bf
r})$ of the condensates which are obtained within the Thomas-Fermi
(TF) approximation. Furthermore, to obtain the densities within the TF
approximation the spatial dependence of the coupling constants is
also neglected by using results of the homogenous system to relate
the density to the chemical potential. It has been shown in the
Refs. \cite{tantar1,lee} that for large $N$ the TF approximation
and spatially independent form of the coupling constant yield
sufficiently accurate results.

We begin the discussion of the results with the values of relevant
parameters from the experiments of Gorlitz et al. \cite{gorlitz} with
$^{23}Na$ atoms. These parameters are
$N=10^{5}$, $\lambda = 26.33$ and $\tilde{a} = 3.8\times
10^{-3}$. We note here that this value of N is consistent
with the TF approximation. These parameters indicate that the condensate
produced in the experiment falls within the Q3D regime.
The numbers obtained by us
for the monopole frequencies of this system are $\Omega_{Q2D} =
2.001$ and $\Omega_{2D} = 2.084$. These results clearly show that
for the above system the Q2D result is very close to the
corresponding Q3D number. However, the 2D model overestimates the
frequency of the monopole mode and this is anticipated as this model is not applicable
to the condensate achieved in the above-mentioned experiment.

Now to study the effect of different models of the coupling
constant and their applicability we choose three different values
of the parameter $\tilde{a}$: $\tilde{a} = 3.8\times 10^{-3}$,
$0.33$ and $2.68$. These values are chosen such a way that the
first, the second and the third numbers fall in the Q3D, the
Q2D and the 2D regimes respectively. In addition to this we also
choose $\lambda = 2\times 10^{5}$ so that the condensate has negligible
length in the z-direction and the motion along this axis is
completely frozen. The results with these parameters are presented
in Table I. Again we can see from Table I that for $\tilde{a} =
3.8\times 10^{-3}$ the numbers predicted by the Q3D and the Q2D
models are very close and the corresponding result from the 2D
model is quite higher than these two numbers. For $\tilde{a} =
0.33$ the numbers obtained by the both Q2D and 2D models are
markedly different from the result of Q3D model. As has been
discussed before for $\tilde{a}=0.33$ along with the large value of $\lambda$
the scattering properties start to get influenced by the
confinement. In this situation it is expected that the Q2D and 2D models will give significantly different numbers in comparison to the corresponding Q3D result.
In the light of our earlier discussion we expect that for $\tilde{a}=0.33$ the coupling constant is better described by the Q2D model. On the other hand, for this value of $\tilde{a}$ the frequency of the monopole mode obtained with the 2D model is higher than that of Q2D model.  As the
the interaction parameter $\tilde{a}$ is further increased to a
value $\tilde{a} = 2.68$ the scattering properties become truly
two dimensional, consequently for this value of $\tilde{a}$ the 2D
model should be able to predict the frequency of the monopole mode
of a two dimensional condensate. In contrast to the case of $\tilde{a}=0.33$,
for  $\tilde{a}=2.68$ the value of the frequency obtained by the
2D model is lower than the corresponding number from the Q2D
model.

Finally, for the sake of completeness we plot in Fig. 1 the
frequencies of the monopole mode obtained with three different
models as a function of the interaction parameter $\tilde{a}$. The
curves are drawn with the anisotropy parameter $\lambda =
5\times10^{5}$ and the number of atoms $N = 10^{5}$. It can be
clearly seen from Fig. 1 that the frequencies of the monopole mode
obtained with the three models of the coupling constant exhibit different trend
with the increase in the interaction parameter $\tilde{a}$. For
example, in contrast to the constant value of the frequency for
the Q3D case the frequencies obtained by employing the Q2D and the
2D models increase as $\tilde{a}$ is increased.
As mentioned before it is only for $\tilde{a}<< 1$ the values of
frequencies obtained by the Q3D and the Q2D models are very close
and the 2D model gives quite different numbers. On the other hand, 
when $\tilde{a}$ exceeds the value $10^{-2}$, the effects of reduced dimensionality start
affecting the scattering properties and both Q2D and 2D models give
different results as compared to the Q3D numbers.
Therefore, we conclude from our results that the collective frequencies of the monopole mode and their behaviour can be used to identify the dimensionality of the
systems as the system evolves from Q3D to a strictly 2D regime. The results of theoretical calculations can also be used to test the validity of different models of the
coupling constant by comparing them with the
experimental values.

In summary, we have calculated the frequencies of collective
oscillations of the Bose-Einstein condensate confined in a flat
pancake-shaped trap. The condensate is tightly trapped along z-axis such
that the motion along this axis is frozen. For such a condensate
depending on the value of the ratio between the size of the
condensate along the z-direction and the s-wave scattering length,
three different regimes can be identified identified. These three different
regimes are described by three different models for the boson-boson
interactions. We have calculated the frequencies of the
of collective oscillations corresponding to these three models. For this
purpose we have used sum-rule approach of many body response
theory along with the ground-state density obtained within the TF
approximation. The main result of this paper is that the different
models for the coupling constant are clearly manifested in the
frequency of the monopole mode and they lead to distinct results
in the region of their applicability. The effect of modification
of the collision properties in two dimension start changing the
monopole frequency when the value of $\tilde{a}$ becomes more
than $10^{-2}$. On the other hand, quadrupole mode is not explicitly dependent on the coupling parameter and thus not affected by the boson-boson interactions.

\section*{Acknowledgement}
It is a pleasure to thank Dr. S. C. Mehendale for helpful discussions and critical reading of the manuscript.

\clearpage
\newpage
\section*{Figure captions}
{\bf Fig.1}Frequencies (in units of $\omega_{\bot}$) of the
monopole mode of $5\times 10^{5}$ $^{23}Na$ atoms confined in a
highly deformed trap with $\lambda = 2\times 10^{5}$ as a function
of the interaction parameter $\tilde{a}$. The solid line
represents results for Q2D case while the corresponding 2D results
are shown by the dashed line and the Q3D results are displayed by
the horizontal line.

\clearpage
\newpage
\begin{table}
\caption{Frequencies of the monopole mode in the units of
$\omega_{\bot}$ for three different values of the
dimensionless interaction parameter $\tilde{a}$ calculated using
Eq. \ref{eqmonopole}for $N = 5\times 10^{5}$ and $\lambda = 2\times 10^{5}$} \tabcolsep=0.5in
\begin{center}
\begin{tabular}{|c|c|c|c|}
\hline
$\tilde{a}$ & $\Omega_{Q3D}$ & $\Omega_{Q2D}$ & $\Omega_{2D}$  \\
\hline
3.8$\times$ 10$^{-3}$& 2.00 & 2.001 & 2.037    \\
0.33 & 2.00 & 2.049 & 2.084    \\
2.68 & 2.00 & 2.292 & 2.182    \\
\hline
\end{tabular}
\end{center}
\end{table}
\clearpage
\newpage
\begin{figure}
\begin{center}
\includegraphics[width = 12cm, height= 15cm]{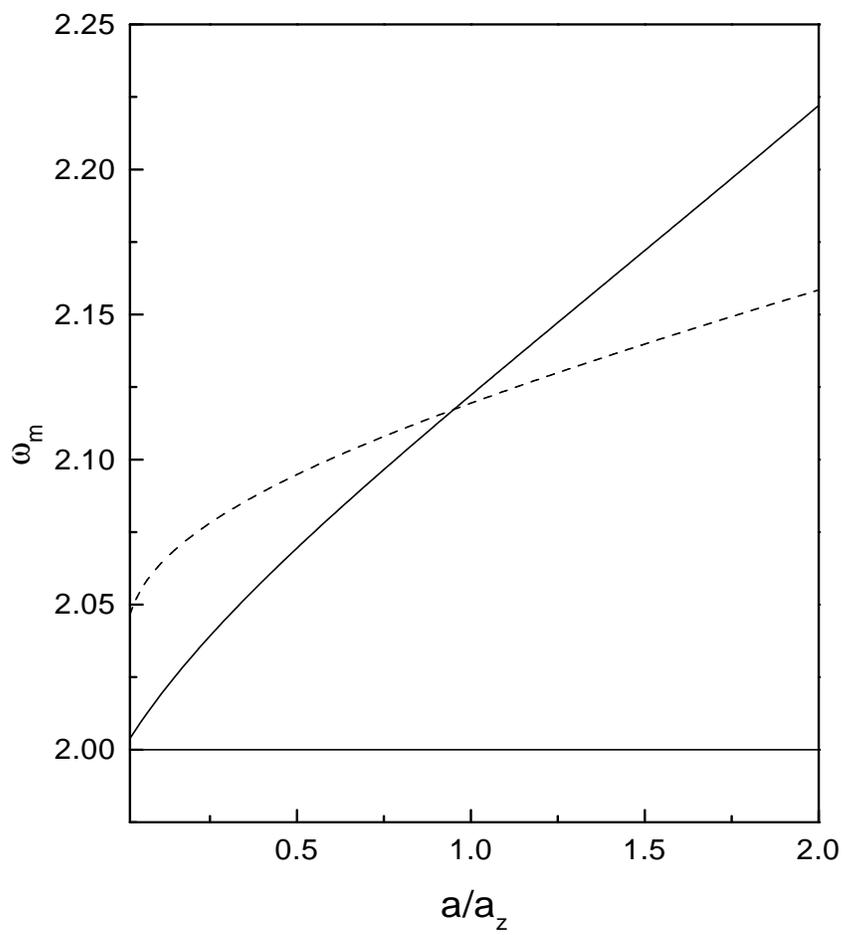}
\caption{Caption rotates along the figure.}
\end{center}
\end{figure}

\end{document}